\documentstyle[12pt]{article}
\begin{document}
\newcommand{\nS}{\mbox{$n_{S}$}}
\newcommand{\nT}{\mbox{$n_{T}$}}
\newcommand{\xT}{\mbox{$x_{T}$}}
\newcommand{\CT}{\mbox{$C^{T}_{2}$}}
\newcommand{\CS}{\mbox{$C^{S}_{2}$}}
\newcommand{\Hpres}{\mbox{$100h\,{\rm km\,sec^{-1}\,Mpc^{-1}}$}}

\begin{center}
{\Large \bf LIMITS ON DIRECT DETECTION OF GRAVITATIONAL WAVES}
\end{center}

\vspace{.2in}
\begin{center}
Rennan Bar-Kana\footnote{email: barkana@arcturus.mit.edu} \\
{\it Department of Physics, MIT
Cambridge, MA 02139, USA.}

\vspace{.2in}
CSR-AT-94-2 \\
Submitted to {\it Physical Review D}
\end{center}

\begin{quote}
{\bf Abstract:}
We compute energy density and strain induced by
a primordial spectrum of gravitational waves on
terrestrial- and space-based detectors (e.g., LIGO)
as constrained by the COBE detection of microwave
background anisotropy.
For the case where the spectrum
is created during inflation, we find new, stricter
upper bounds on the induced strain,  making detection
unlikely.  However, detectors might be useful for
discovering (or ruling out) exotic, non-inflationary
sources.
\end{quote}

\begin{center}
\vspace{.2in}
PACS numbers~~~98.80.Cq, 04.30.-w, 04.80.Nn\\
\end{center}

\vspace{.3in}

Events in the early universe may have left a primordial spectrum of
gravitational waves. Detecting these elusive remnants would not
only establish the existence of this as-yet-unverified prediction of
General Relativity, but it would also provide a new critical test for
all proposed scenarios of the evolution of the early universe. In
particular, a basic feature of  the inflationary model of the universe is
the prediction of a relic spectrum of gravitational  waves
\cite{AbbWs}, whose
detection would lend strong support to the theory.

Such a detection might occur in three possible ways. Gravitational
waves distort the Cosmic  Microwave Background (CMB) through
the Sachs-Wolfe effect \cite{ScsWlf}, thus raising the possibility
that  some or even most of the temperature variations observed by
COBE \cite{COBE} are due to gravitational waves \cite{Davis}.
Another possibility for detecting tensor fluctuations indirectly is
through their effect on the polarization of the microwave background
\cite{Polariz,Polariz2}. In this Brief Report, we consider the possibility
of direct observation of the primordial  gravitational background, in
a laser interferometer gravitational wave detector \cite{LIGO}
or through its effect on pulsar timing \cite{Pulsar}.

Gravitational waves may be produced by many sources. At  the
Planck time, quantum  fluctuations in the metric are significant and
they produce  gravitons. Phase transitions in the  universe may lead
to topological defects such as cosmic strings, which generate
gravitational waves. A period of inflation leaves behind
gravitational waves. We first  consider the case of gravitational
waves produced by inflation, and discuss revised predictions for
the strain induced in direct detection. Independent analyses have
been made by \cite{Sahni},\cite{Sordp},\cite{Liddle},
\cite{NewTurn}. We then broaden the
discussion to consider more general spectra, and the limits that may
be placed  on their existence.

In inflation, gravitational waves are produced in conjunction with
density fluctuations.  Inflation is a proposed solution to the
homogeneity, flatness, and  monopole problems that  are
unexplained in the standard Big Bang model. The essential  idea of
inflation is that the  universe underwent a period of extraordinarily
rapid expansion $10^{-35}$ seconds or so after the Big Bang \cite{Infl}.
Both density perturbations \cite{scalar} and gravitational
waves \cite{AbbWs} are produced as
a  result of the rapid stretching of quantum fluctuations occurring
during inflation, in the  inflaton field and graviton field respectively.
After inflation, both fluctuations span a broad-band spectrum,
ranging from the scale of the present horizon down to microphysical
scales. The variation of amplitude with scale can be approximated as
a power law. The index for scalar (density)  fluctuations
we denote \nS, and that for tensor (gravitational wave) fluctuations
\nT, where we choose a convention in which a strictly
scale-invariant spectrum corresponds to $\nS =  1$ or
$\nT = 0$ \cite{Davis}. The broad spectrum makes it
possible to detect gravitational waves  using CMB anisotropy on
cosmological scales, pulsar timing measurements on  astrophysical
scales, and laser interferometer detectors on terrestrial scales.

If quadrupole temperature fluctuations in the CMB consist of long
wavelengths  contributions of gravitational waves (\CT ) and
energy-density perturbations (\CS ), then the COBE measurement
of large-angular CMB anisotropy fixes the sum $\CT +\CS$.  This
leaves the ratio undetermined. Naive gravitational wave
limits for inflation had assumed $\CS = 0$,
and scale invariance $\nT
= 0$. Using COBE this leads to the $\nT =  0$ prediction (dotted
curve) shown in Fig. 1. Here we consider the implications of a
recent refinement of the inflationary predictions.\cite{Davis,Liddle,NewTurn}
Namely, inflation
does not predict precisely scale-invariant spectra. Rather, models of
inflation give various values of \nT\  and \nS\ but  with
\begin{equation}
\label{eq:nT}
\nT \approx \nS - 1 .
\end{equation}
For each \nT\ the fraction of the gravitational waves contribution to
the CMB quadrupole anisotropy is predicted:
\begin{equation}
r = \frac{\CT}{\CS}  \approx -7\nT
\label{eq:xT}
\end{equation}
\cite{1stOrder}
Eq.~\ref{eq:xT}  is an accurate approximation
for generic inflationary potentials.
Exceptions require
fine-tuning of parameters or initial
conditions, beyond that which is strictly necessary
for inflation.  Examples include cosine potentials  (`natural inflation')
 or potentials in any inflationary model  in which an extremum
  is encountered near the end of inflation.
  For these exceptional models, the gravitational wave amplitude is
  less than the generic case and  our calculations  are
overestimates.\cite{InfPot,InfModels}

The fraction $\xT = \frac{\CT}{\CT+\CS}$ determines
the gravitational wave amplitude at long
($O(H^{-1}_{0})$) wavelengths and \nT\ determines the relative
amplitude on smaller wavelengths. Eq.~\ref{eq:xT} modifies the
predictions for gravitational wave detectors. A strictly scale-invariant
spectrum ($\nT = 0$) is now forced towards $\xT = 0$. To obtain
an appreciable amplitude, \xT\ must be $> 0$; however, this only
occurs if $\nT < 0$, reducing the relative amplitude on smaller
scales below that of the scale-invariant case with the same \xT.
Hence, Eq.~\ref{eq:xT} reduces the expected strain in direct
detectors for gravitational waves coming from inflation.

It is also possible that there is a non-inflationary energy density
spectrum of gravitational waves from another source, with some
index \nT. Regardless of the source, if such a spectrum contributed
significantly to the COBE anisotropy (long wavelengths), then a
high enough \nT\ would make the shorter wavelengths directly
detectable by gravitational wave detectors. Such a detection would
indeed cause excitement, since there is no established mechanism
that generates such a spectrum. Conversely, a lack of detection at the
sensitivities of proposed detectors would serve to place an upper
limit on such exotic spectra.

Fluctuations in the metric produce temperature anisotropies in the
CMB through the Sachs-Wolfe effect. These temperature
fluctuations $\frac{\Delta T}{T}$ can be written in terms of
spherical harmonics. If
\begin{equation}
\frac{\Delta T}{T}=\sum_{l,m} a_{lm} Y_{lm}(\theta,\phi)
\end{equation}
then the quadrupole is
\begin{equation}
C_{2}= \langle |a_{2m}|^{2}\rangle.
\end{equation}
The rms quadrupole amplitude measured by COBE \cite{COBE} is
\begin{equation}
Q_{T}^{2}=\frac{5}{4 \pi}C_{2}.
\label{eq:COBE}
\end{equation}
Inflation predicts that the tensor and scalar contributions are
independent, $C_{2}= \CT+\CS$.

The quadrupole \CT\ is produced by graviton modes of long
wavelengths $\lambda$.  Here $\lambda$ denotes the comoving
wavelength, which is the same as the present  physical wavelength
$\lambda ^{0}_{phys}$ if we set the present scale factor $R_{0}=
1$. We also assume a flat $\Omega=1$ universe and set $h=0.75$,
where $H_{0}=\Hpres$. For a
gravitational wave spectrum with index \nT, the energy  density of a
given mode outside the horizon is proportional to $\nu ^{n_{T}}$,
expressed  in terms of the frequency $\nu = \frac{c}{\lambda}$.
Horizon crossing for a given mode at  time $t$ is defined by
$\lambda^{t}_{phys}=2 H^{-1}_{t}$ (in particular $\nu _{0} =
\frac{c}{2H^{-1}_{0}}$). At horizon crossing, $\Omega
_{g}(\nu) = \frac{1}{\epsilon_{cr}} \nu
\frac{d\epsilon_{g}}{d\nu}\sim \nu ^{n_{T}}$  ($\epsilon_{g}$ is
the energy density of the gravitons, $\epsilon_{cr}$ is the critical
value  of the energy density) \cite{Sahni}. Once the mode crosses
inside, its energy density  redshifts $\propto R^{-4}$, as for a
relativistic species. During the radiation dominated  epoch,
$\epsilon_{cr}$ is also $\propto R^{-4}$ and therefore $\Omega
_{g}(\nu)$ stays constant. During the matter dominated epoch
$\epsilon_{cr} \propto R^{-3}$ and so  $\Omega _{g}(\nu) \propto
R^{-1}$. A more careful treatment of the transition from  radiation
to matter domination yields, in terms of the quadrupole anisotropy, a
spectrum at the present of \cite{NewTurn,Sahni,Sordp}
\begin{equation}
\Omega _{g}(\nu)=\frac{\xT\, C_{2}}{15\, g(\nT)}\, T(\nu)\,
(\frac{\nu}{\nu_{0}})^{n_{T}- 2}
\label{eq:Omega}
\end{equation}
where $T(\nu) \approx  1+3.49 R_{EQ}^{1/2}\frac{\nu}{\nu_{0}}+16.9
R_{EQ}  (\frac{\nu}{\nu_{0}})^{2}$, $g(\nT) \approx exp[1.3 \nT]$,
\cite{AbbWs} and  $R_{EQ}=4.18 \times 10^{-5}
h^{-2}$.  \cite{1stOrder}

In Fig. 1, the spectrum of gravitational waves from inflation has
been calculated as a  function of \nT\  using Eq.~\ref{eq:xT},
Eq.~\ref{eq:COBE} and Eq.~\ref{eq:Omega}.  We describe the
predictions in terms of the dimensionless strain $h_{g}^{2}(\nu) =
24 \Omega _{g}(\nu)/(\frac{\nu}{\nu_{0}})^{2}$. A
comparison of the strain  curves with the projected sensitivities of
LIGO\cite{LIGOII}
(the Laser Interferometer Gravitational  Wave Observatory) and
LAGOS (the Laser Gravitational Wave Observatory in Space),
and current experimental limits from Pulsar timing \cite{LIGO}
shows that the inflationary  maximum ($\nT = -0.02$) lies $0.5$
of an order of magnitude below the current estimate of the
sensitivity of LAGOS and $1.5$ orders of magnitude below
the estimate for LIGO-Advanced Detectors. Note that all of
the  inflationary predictions, based on Eq.~\ref{eq:xT} (solid
curves) lie below the naive limit for $\nT  = 0$ assuming $\xT = 1$
(dotted curve). The maximum strain at LAGOS and LIGO
wavelengths is found to occur for $\nT  = -0.02$, a factor of
$4$ below the naive limit.

For a general, non-inflationary spectrum, we must replace
Eq.~\ref{eq:xT} by  some other assumption.
If we assume $\xT=1$ then we find that LIGO I can be used
to detect spectra with $\nT > 0.3$
($\nT < 0.1$ for LIGO II).
LAGOS would be sensitive to spectra
down to the scale-invariant
$\nT = 0$. (See Fig. 2) If we assume $\xT=1\%$
instead we obtain $\nT > 0.4$ for LIGO I,  $\nT > 0.2$ for LIGO
II, and $\nT > 0.1$ for LAGOS. The best experimental limits that
we can currently achieve are derived from Pulsar timing
measurements. For $\xT=1$,  we get $\nT < 0.7$, and for
$\xT=1\%$, we get $\nT < 0.9$.

An upper bound on \nT\ for $\nT > 0$ can presently be
obtained by considering  the contribution of the gravitational waves
to the total $\Omega$ of the universe, and using  the constraint
$\Omega < 2$. If the spectrum is produced at time $t$ with
corresponding  $\nu_{t}$  then
\begin{equation}
\Omega = \frac{\epsilon_{g}}{\epsilon_{cr}}=\int
_{\nu_{0}}^{\nu_{t}} \frac{\Omega_{g} (\nu)}{\nu} d\nu.
\end{equation}
Assuming $\nu_{t} > 10^{4}$ (corresponds to LIGO) we
obtain the following limits (see Fig. 3): $\nT \leq 0.7$ for  $\xT=1$,
$\nT \leq 0.8$ for $\xT=1\%$. These limits on gravitational waves
are the best now available, but it appears that projected experiments
will soon yield much stricter limits. Note that our limits do not apply
for spectra which are not of power-law form. For example,
gravitational waves produced by bubble collisions at the end of
inflation are peaked over a narrow range of frequencies
\cite{foigw}.

\vspace{.1in}

I thank Paul J. Steinhardt for valuable discussions and comments,
and the University Scholar's Program of the
University of Pennsylvania, which provided funding for this work.

\newpage

%%%%%Insert Figure gw1.ps
\begin{figure}
\vspace{3.5truein}
\caption{The dimensionless strain of gravitational waves from
inflation (solid curves) vs. the present physical frequency, for $\nT
=0, -0.02,-0.15$ and $-0.30$. The scale-invariant curve assuming
$\xT = 1$ (dotted curve) is also shown. The projected sensitivities
of the LIGO I, II and LAGOS detectors are shown (shading) for
comparison.}
   \includegraphics{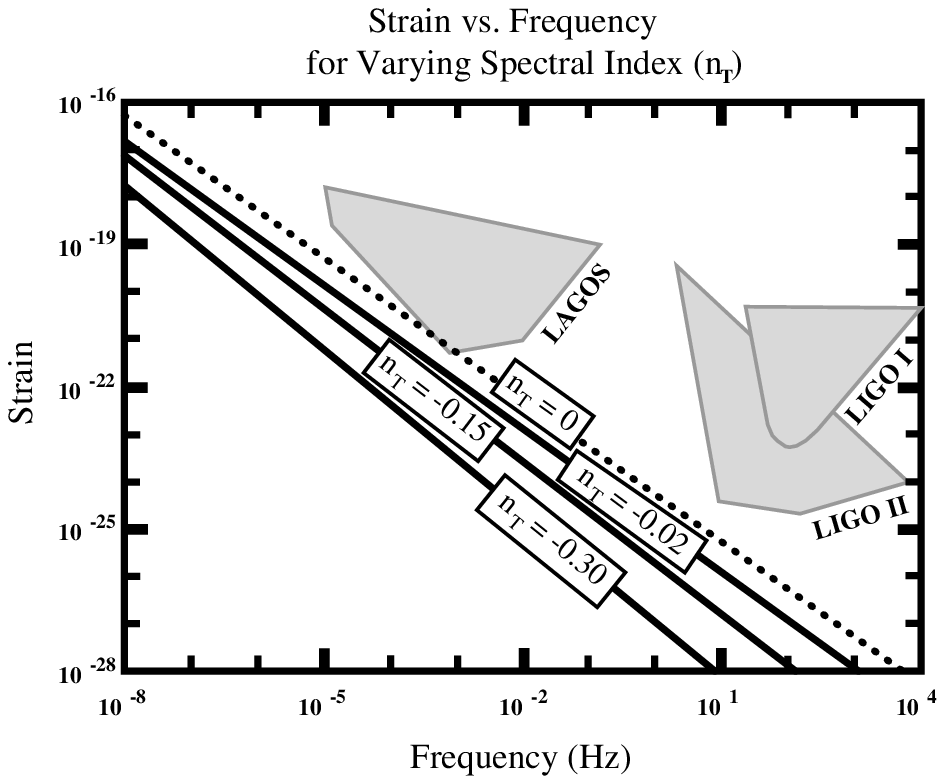}
\end{figure}
%%%%%%%%%%%%%%%

%%%%%Insert Figure gw2.ps
\begin{figure}
\vspace{3.5truein}
\caption{The dimensionless strain of non-inflationary gravitational
waves all assuming $\xT = 1$ (solid curves) vs. the present physical
frequency, for $\nT = 0.1,0.3$ and $0.7$. The scale-invariant curve
assuming $\xT = 1$ (dotted curve) is also shown. The projected
sensitivities of the LIGO I, II and LAGOS detectors as well as
currents limits from Pulsar timing measurements are shown
(shading) for comparison.}
   \includegraphics{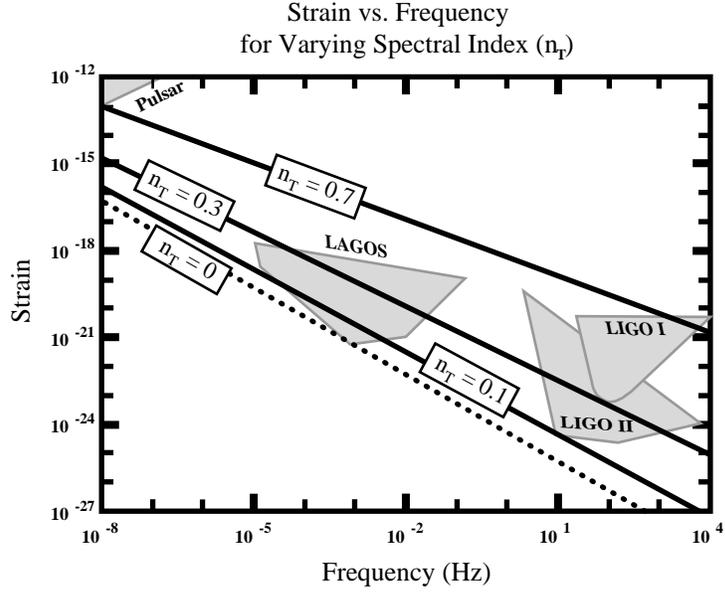}
\end{figure}
%%%%%%%%%%%%%%%

%%%%%Insert Figure gw3_2.ps
\begin{figure}
\vspace{3truein}
\caption{The spectral energy density (in units of the critical energy
density) of non-inflationary gravitational waves assuming $\xT = 1$
vs. the present physical frequency, for $\nT = 0,0.3$ and $0.7$.}
   \includegraphics{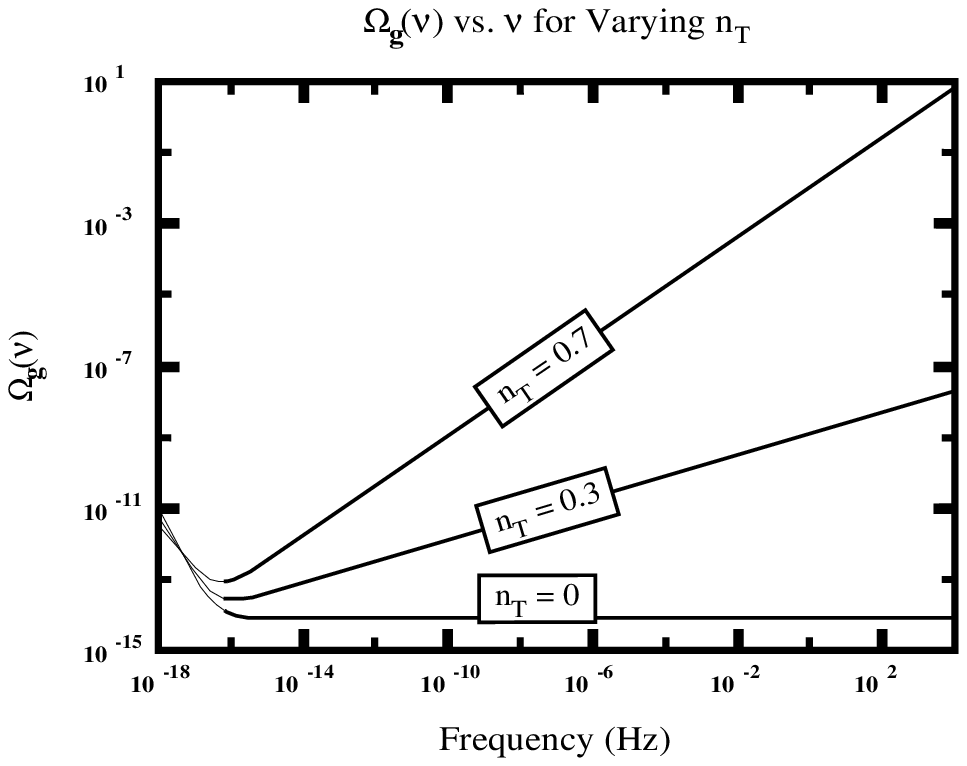}
\end{figure}
%%%%%%%%%%%%%%%


\begin{thebibliography}{gwaves}
\bibitem{1stOrder} For the inflationary calculation, Eq.~\ref{eq:xT}
is only correct to 1st order in \nT, and so all factors, including
$g(\nT)$ in Eq.~\ref{eq:Omega}, will be entered to lowest order
in \nT\ in the final result.
\bibitem{InfPot} {We assume that the inflationary potential
is a continuous function of the field and monotonically decreasing
 over the last  60 e-folds.}
\end{thebibliography}
\end{document}